\documentstyle[11pt,aaspp4]{article}

\def\wisk#1{\ifmmode{#1}\else{$#1$}\fi}

\def\kms{~km s$^{-1}$}
\def\coonezero{CO($1\rightarrow 0$)}
\def\coa32{CO($3\rightarrow 2$)}
\def\13coa32{$^{13}$CO($3\rightarrow 2$)}
\def\cob43{CO($4\rightarrow 3$)}
\def\coc54{CO($5\rightarrow 4$)}
\def\cod76{CO($7\rightarrow 6$)}

\def\hcna10{HCN($1\rightarrow 0$)}
\def\hcnb43{HCN($4\rightarrow 3$)}
\def\car10{CI($^3P_1\rightarrow {^3P_0}$)} 
\def\carb21{CI($^3P_2\rightarrow {^3P_1}$)} 
\def\gtrapprox{\;\lower 0.5ex\hbox{$\buildrel >
    \over \sim\ $}}             
\def\lessapprox{\;\lower 0.5ex\hbox{$\buildrel < \over \sim\ $}}
\def\etal{et al.\ }
\def\nco{\wisk{N_{\rm CO}}}
\def\dv{\wisk{\Delta V}}
\def\htwo{\wisk{\rm H_2}}
\def\psqcm{\ifmmode {\>{\rm cm}^{-2}}\else {cm$^{-2}$}\fi}
\def\pcubcm{\ifmmode {\>{\rm cm}^{-3}}\else {cm$^{-3}$}\fi}
\def\be{\begin{equation}}
\def\ee{\end{equation}}
\def\bea{\begin{eqnarray}}
\def\eea{\end{eqnarray}}
\def\rpcsq{\ifmmode {r_{\rm pc}^2}\else {$r_{\rm pc}^2$}\fi}
\def\rpc{\ifmmode {r_{\rm pc}}\else {$r_{\rm pc}$}\fi}
\def\fabs{\ifmmode {f_{\rm abs}}\else {$f_{\rm abs}$}\fi}
\def\msol{\ifmmode {\>M_\odot}\else {$M_\odot$}\fi}
\def\lsol{\ifmmode {\>L_\odot}\else {$L_\odot$}\fi}

\begin{document}

\title{Multiple CO Transitions, CI, and HCN from the Cloverleaf Quasar}

\author{Richard Barvainis}
\affil{MIT Haystack Observatory, Westford, MA 01886 USA\footnote
{Radio Astronomy at the Haystack Observatory of the 
Northeast Radio Observatory Corporation (NEROC) is supported by a grant 
from the National Science Foundation}}

\author{Philip Maloney}
\affil{CASA, University of Colorado, Boulder, CO 80309-0389 USA}

\author{Robert Antonucci}
\affil{UC Santa Barbara, Physics Department, Santa Barbara, CA 93106 USA}

\and
\author{Danielle Alloin}
\affil{Service d'Astrophysique, L'Orme des Merisiers, 
CE Saclay, 91191, Gif-Sur-Yvette, Cedex, France}







\begin{abstract}
New millimeter-wavelength emission lines are reported for the
Cloverleaf (H1413+117), a gravitationally lensed quasar at redshift 2.56.  
High signal-to-noise ratio spectra of four lines in CO ($J = 3\rightarrow
2,~4\rightarrow 3, ~5\rightarrow 4,~{\rm and}~7\rightarrow 6$) have
been obtained and are modelled using an escape probability formalism.
The line brightness temperatures are flat or rising between \coa32 and
\cob43 and then drop to the \coc54\ and \cod76\ lines; this
falloff suggests that the optical depths in the CO lines are 
modest ($\tau_{4-3} \lesssim 3$), and that the gas is relatively
warm ($T \gtrapprox 100$ K) and dense ($n_{\rm H_2} \gtrapprox 3\times
10^3$ cm$^{-3}$).  

The neutral carbon fine structure line \car10 was detected with
moderate SNR on two separate occasions, and appears secure. The
HCN($4\rightarrow 3$) line is somewhat more tentative, as it was
observed only once and detected with moderate SNR. However, its
brightness ratio relative to the \cob43 line, $L'({\rm
HCN})/L'({\rm CO})\approx 0.25$, is very similar to the ratio seen in
luminous infrared galaxies. This probably reflects the gas
fraction maintained at high densities by radiation pressure from
the active nucleus; alternatively, the HCN transition may be excited
through absorption of infrared radiation at $14\mu$m.

The detectability of the Cloverleaf in molecular and neutral atomic
transitions is largely due to high emissivity of the gas and 
magnification from gravitational
lensing, rather than an extremely large mass of gas.

\end{abstract}


\section{Introduction}

There are few observational windows on galaxies at high redshift, and
each new one that opens up is invaluable.  It is now possible to
directly study interstellar gas in significant quantity in three 
systems at $z > 2$: IRAS F10214+4724 ($z = 2.28$), the Cloverleaf
quasar H1413+117 ($z = 2.56$), and the quasar BR1202--0725 ($z
= 4.69$).  F10214+4724 was first identified as a luminous infrared
galaxy (Rowan-Robinson et al.\ 1991), and has been detected
in three transitions of carbon monoxide (Brown \& Vanden Bout 1992a;
Solomon, Downes, \& Radford 1992a). It is now known both that this object
is gravitationally lensed (e.g. Broadhurst and Leh\'ar 1995), and that it
harbors an obscured quasar nucleus seen in polarized (reflected) light
(Goodrich et al.\ 1996). BR1202-0725, the third highest redshift quasar
currently known, has recently been detected in
CO by Ohta et al.\ (1996) and Omont et al.\ (1996). 
The Cloverleaf quasar, the subject of this paper, is a
QSO with classical broad emission lines plus broad absorption troughs
in the optical, with an image that is lensed into four spots separated
by about $1''$.

Based on its strong submillimeter continuum emission, indicative of a
large mass of cool dust (Barvainis, Antonucci, \& Coleman 1992), the
Cloverleaf presented a good candidate for the study of
molecular gas at high redshift. Indeed, CO($3\rightarrow 2$) line
emission, redshifted to 97 GHz, was searched for and readily detected
(Barvainis et al.\ 1994; Wilner, Zhao \& Ho 1995). The close
similarities in both infrared/submillimeter continuum and CO emission
properties between the Cloverleaf and F10214+4724, and the fact that
both have powerful active nuclei, have given rise to models uniting
these two objects through orientation/obscuration effects (e.g.,
\cite{bar95}; Granato, Danese, \& Franceschini 1996), and have
bolstered evolutionary scenarios linking luminous infrared galaxies
and quasars in general (e.g., Sanders et al.\ 1988; Wills et al.\ 1992).

To provide more information on the state of the interstellar
medium in the host galaxy of the Cloverleaf, we have obtained
detections in three new transitions of CO, and have made measurements
in two transitions of neutral carbon and one of HCN.  We
present these results, together with models for the emitting gas,
below.

\section{Observations and Results}

The observations were carried out using the IRAM 30m telescope
on Pico Veleta, Spain, during observing runs in 1994 June (two nights)
and 1995 May (seven nights), and with the Plateau de Bure Interferometer
(PdBI),
near Grenoble, France, during 1996.  At the 30m the weather was mainly clear
during both runs, but occasional high winds during the 1995 run
resulted in some downtime. Spectral data were obtained in the 3mm,
2mm, and 1mm bands simultaneously, using SIS mixer receivers and
filterbank or acousto-optical backends of 512 MHz bandwidth.  System 
temperatures derived
using the chopper wheel method were typically 250 K, 200 K, and 400 K
in the 3mm, 2mm, and 1mm bands respectively.  Double position
switching was employed to provide flat baselines, with subreflector
nutation at a rate of 0.5 Hz.

Clear detections were obtained in four CO transitions: $J=3\rightarrow
2,~4\rightarrow 3, ~5\rightarrow 4,~{\rm and}~7\rightarrow 6$.  Peak
antenna temperatures were in the range 4--8 mK.  
$^{13}$CO($3\rightarrow 2)$ was searched for but not detected.  
A similar line flux and velocity width for the $J=3\rightarrow 2$ line
was obtained by Wilner \etal (1995).  
The \cod76 single-dish measurement presents a problem, because
of the limited available bandwidth of 512 MHz.  Corresponding to 680\kms\ 
at 226 GHz, this is not wide enough to provide a zero-level baseline -- 
even with subreflector nutation the zero level in the raw spectrum
cannot be trusted.  We will therefore instead use here the PdBI 
observations of \cod76 (for details, see Alloin et al.\ 1997).  
These also suffer from the same limited bandwidth, but for  
interferometer observations the zero level should be accurate.  
To derive the total 
intensity (Jy\kms ) of the line for comparison with the other CO 
transitions it is necessary to fit the line with a gaussian, but the 
line width is poorly constrained since the line wings are cut off at 
the band edges.  Fixing the linewidth to the mean of the other three
CO transitions, $\Delta V_{\rm FWHM} = 376$ \kms , 
we obtain $I_{\rm PdBI} = 43.7 \pm 2.2$ Jy\kms\ (the error is statistical
only, and does not include 10\% calibration uncertainty).

The spectra are shown in Figure 1, in units of flux density.  Table 1
gives line parameters and upper limits as derived from Gaussian
profile fits to the 30m spectra [except for \cod76, where the line
parameters are estimated as described above]. The line luminosities
are given in two ways: $L$, in units of solar luminosities, and $L'$
in units of K km s$^{-1}$ pc$^2$ (see Solomon, Downes, \& Radford
1992a for definitions and formulae for computation).  $L'$ is
proportional to line brightness (Rayleigh-Jeans) temperature
integrated over the area of the source, and assuming that the CO lines
arise in the same general volume this quantity can be used to form
line brightness temperature ratios to constrain physical conditions in
the gas (see below).

\placefigure{fig1}

Observations were also made of the neutral carbon fine structure lines
\car10\ and \carb21\ (rest frequencies 492.161 GHz and 809.343 GHz,
respectively). A detection in the former made in the 1994 run
repeated with reasonable consistency in 1995, and the average
of all good scans from both years is shown in Figure 1. Confidence in
this detection is enhanced by the similarity in redshift and line shape
to the CO lines (Table 1). Also, the line repeated after
a shift in observing frequency which placed it in a different part of
the filterbank. In \carb21, only an upper limit was obtained (Table 1).

Finally, we observed \hcnb43\ during the 1995 run. We report a
tentative detection of a line with peak flux density of 5.5 mJy 
(Figure 1). The total integration time for this observation
was about 30 hours. Although this line is near the limit
of detectability for the 30m telescope, as in the case of \car10 we
are encouraged by the similarity of the line parameters to those of CO.

Some comments on the uncertainties in the observed line ratios
are in order. In addition to the formal errors from noise
in the measurements, it is necessary to include the systematic
errors. Conversion from antenna temperature in Kelvins to flux
density in Janskys was done using standard calibration factors for
the 30m telescope derived from observations of planets and other
standard sources.  The errors in these factors  are estimated to be 
$\pm 10\%$ 
for the $3\rightarrow 2$ line and $\pm 15\%$
for the $4\rightarrow 3$ and $5\rightarrow 4$ lines (C. Kramer, 
private communication). For the PdBI observations of \cod76\, the 
calibration uncertainty is estimated at 10\%.  Including the
systematic and statistical errors, the CO line Rayleigh-Jeans
brightness temperature ratios relative to the strongest line, 
CO($4\rightarrow 3$),
are then:
$(3-2)/(4-3)=0.83\pm 0.16$, $(5-4)/(4-3)=0.73\pm 0.16$;
$(7-6)/(4-3)=0.68\pm 0.13$.   These ratios are computed from the 
$L'$ values given in Table 1.

\placetable{tbl-1}

\section{Interpretation}

\subsection{The CO lines}

The CO lines are characterized by a constant or increasing brightness
temperature in going from $J=3\rightarrow 2$ to $J=4\rightarrow 3$,
followed by a decrease in the higher-$J$ transitions.  We cannot rule
out the possibility that the brightness temperatures are the same for 
all transitions, since the line ratios differ 
from predicted thermal R-J values by only $1.5-2\sigma$ in each case.
In the case of thermal line ratios we would be able to say little about 
the column density or temperature of the gas.
However, below we will assume that our measured values represent the
true line brightnesses, and use them to model physical conditions in 
the gas.  This exercise 
suggests that the line optical depths are modest, with $\tau\sim$ a few 
or less.


In modeling the CO line ratios, we calculate the excitation of the
CO molecules using a non-LTE scheme which employs mean escape
probabilities to include the effects of radiative trapping on
the excitation of the CO (cf. \cite{aa91}); the results depend
on the density $n$ of \htwo , the gas kinetic temperature $T$, and the CO
column density per unit linewidth \nco/\dv. We note in passing that
in non-LTE models using the essentially equivalent LVG approximation,
the ratio of CO abundance to velocity gradient -- which fixes the
optical depths -- is often set to a value appropriate to Galactic
molecular clouds (e.g., Solomon, Downes, \& Radford 1992a), which is 
not {\it a priori} justifiable when the physical conditions in the gas 
may be radically different.

In Figure 2 we show the allowed ranges 
in $n$ and \nco/\dv\ for three temperatures, $T=60$, 100, and 300 K;
the allowed ranges are the hatched regions within the curves. A
temperature of 60 K is the minimum acceptable temperature.  We have
also imposed the constraint that the brightness temperature of the
\cod76\ line must be at least 40 K.\footnote{Using the observed {\it
lensed} CO source size in the \cod76\ line (Alloin et al.\ 
1997) indicates a source radius $r\lesssim 0.48 h^{-1} m_{\rm C}^{-1}$ kpc
(assuming $q_0=1/2$), where $m_{\rm C} \approx 0.2m$ is the magnification of 
CO in image C of the quad ($m$ is the total magnification for all four
images). The exact relation between $m_{\rm C}$ and the estimate
of the true source radius $r$ is model dependent --  here we have assumed that
lensing stretches the source predominantly in one dimension; see 
Burke et al (1991) for an illustration of distortions expected in 
a quad lens.  Given these assumptions, 
the minimum brightness temperature in the \cod76\
line (that is, for unit areal filling factor) is $T_R (7-6)\gtrapprox
40 m_{\rm C}$ K, independent of $h$.}  This
constraint can in principle eliminate a number of models which satisfy
the line {\it 
ratios}. For the high temperature models, the allowed range in density
is a factor of $3-10$, while for $T=60$ K any density greater
than $n\approx 8\times 10^4\pcubcm$ is allowed. The acceptable range
of \nco/\dv\ increases from about a factor of two ($T=60$ K) to a
factor of seven ($T=300$ K). 

At $T=60$ K, the line ratio
constraints can be satisfied for thermalized level populations (which
is why there is no upper limit to the acceptable density), provided
that the optical depth is not too large ($\tau_{3-2}\lessapprox 2$):
otherwise the line ratios would simply be the ratio of Rayleigh-Jeans
brightness temperatures for $T=60$ K, 
which is too large compared to the observed line ratios.\footnote{ 
For example, the R-J brightness temperature ratio for optically thick 
and thermalized lines at 60 K is $0.89$ for (7-6)/(4-3), compared with 
the observed value of $0.68\pm 0.13$. At 300 K, the thermal R-J ratio
is 0.98.}
Thus the ratio of the optical depth terms ($1-e^{-\tau}$) must be less 
than one, which requires that the optical depths are not much greater 
than unity and that $\tau_{7-6}$ is less than $\tau_{3-2}$. The
sharp cutoff in $N_{\rm CO}/\Delta V$ at the high end is the result of
this optical depth restriction; the boundary is nearly vertical
because, except at the lowest densities, the rotational levels are
thermalized through $J=7$ for all the densities. 

At high temperatures the line ratios are matched by somewhat different
physics. Thermalized levels are unlikely, as the high$-J$
lines would then be too strong relative to $J=4\rightarrow 3$ no
matter what the optical depths. The observed fall-off to the higher
$J$ lines can still be matched in this case, provided that the level
populations are subthermal; in particular, the excitation temperatures
$T_{\rm ex}$ of the $J=5$ and $J=7$ levels must be well below
that of $J=4$. This restricts the density from being too large, and in
fact there is only a narrow range (at a fixed temperature) at which
the line ratio constraints are satisfied. In addition, the optical
depths are also restricted from being too large, as otherwise
radiative trapping would inhibit the required decrease in the
excitation temperature. It is really the drop in $T_{\rm ex}$ that is
crucial, as the optical depths may actually {\it increase} from the
$3\rightarrow 2$ line to the $7\rightarrow 6$ line;
the drop-off in line flux is then due to the fact that $kT_{\rm
ex}/E_J$ is less than $\sim$ unity for the $J=7$ level, while $T_{\rm
ex}(J=3)$ is non-negligibly larger than $T_{\rm ex}(J=7)$. There is a
systematic trend in the line ratios; the model values approach the
maximum allowed observational values with increasing $N_{\rm CO}/\Delta V$.

The maximum optical depth in the \cob43\ line is generally
$\sim 3$.  The relatively low optical depths are mandated by the low 
value of the $(5-4)/(4-3)$ and $(7-6)/(4-3)$ ratio, as outlined above.
This result is unusual in comparison to Galactic molecular clouds,
where the optical depth in the CO $J=1\rightarrow 0$ line (the most
frequently observed transition) is generally inferred to be
large. However, modest optical depths in the CO rotational
transitions appear to be common in starburst galactic nuclei
(\cite{aa95}), for the same reason as in our models of the CO emission
from the Cloverleaf: the elevated gas temperatures ($T\gtrapprox 60$
K) distribute the molecules over a much larger number of rotational
levels.
In addition, since the Cloverleaf is at a redshift $z =
2.558$, it is possible that the gas-phase metallicity is considerably
below the Solar neighborhood value, which will affect the CO abundance
and emission (e.g. Maloney \& Wolfire 1996). A clear prediction of our
single-component CO models is that the \coonezero\ line should be weak
compared to the \coa32\ line, with a brightness temperature about
one-third that of the $3\rightarrow 2$ line; this is a consequence of the
increase in optical depth with $J$ up to $J\approx 4$ or 5.

\subsection{The CI lines}
Our upper limit on the intensity of the CI 809 GHz line gives an upper
limit to the \carb21/\car10\ ratio of $\sim 3.9$ (here using luminosities
$L$ expressed in $L_{\sun}$).  We have calculated
the expected ratio of these lines as a function of density,
$T$, and $N_C/\Delta V$, as
for the CO lines. The upper limit to the 809/492 ratio is only
significant for gas temperatures $T\gtrapprox 100$ K, where it
restricts the gas densities to be $n\lessapprox 10^4$ cm$^{-3}$. Thus in
principle, the CI emission can arise from the same regions which
produce the CO emission, as delineated in Figure 2. 

The observed ratio of luminosities $L$ in the CI 492 GHz and \coa32\
lines is approximately 0.5 (Table 1).  In order to match this ratio,
the column density per unit linewidth of neutral carbon must be
$N_C/\dv\gtrapprox 3\times 10^{17}$ cm$^{-2}$ $({\rm
km\;s^{-1}})^{-1}$.  Therefore, 
if the CO and CI emission arise from the same regions,  
the neutral carbon abundance is comparable to or a few times larger 
than that of CO.  This unusually high ratio of C to CO is a 
reflection of the rather low ratio of $N_{\rm CO}/\Delta V$ required
to match to CO line ratios.

We can compare the CI/CO ratio for the
Cloverleaf with the {\it COBE} results for the Milky Way (\cite{wri91}),
and extragalactic detections of the 492 GHz line in IC 342 by
\cite{but92}, and in M82 by \cite{sch93} and
\cite{whi94}\footnote{Brown \& Vanden Bout (1992b) have reported detection 
of CI in F10214+4724, but at a substantially different velocity from 
\coa32 . The emission in the two species is therefore unlikely to be 
cospatial, so we do not cite a CI/CO ratio for this object here.}. 
Relative to the CO $J=2\rightarrow 1$ line, the flux in
the 492 GHz CI line is $1.4\pm 0.4$ (IC 342) and $2.3\pm 0.6$ (Milky
Way). If the CO lines are optically thick, the ratio of the CI line to
the CO transitions will scale as $\nu^{-3}$ in the high-temperature
limit, so that relative to the \coa32\ line these flux ratios are 0.4
(IC 342) and 0.7 (Milky Way), similar to the value of 0.5 seen in the
Cloverleaf. From \cite{sch93} and \cite{whi94}, the ratio in M82 is
0.7. \cite{but92} conclude that the ratio
seen in IC 342 and the 
Milky Way is consistent with the bulk of the emission arising in
photodissociation regions (PDRs), although \cite{sch93} and
\cite{whi94} argue that the ratio is too large compared to standard
PDR models; the CI intensity can be enhanced if the PDRs are clumpy or
the cosmic-ray flux is very high. Although the CI and CO emission in
the Cloverleaf probably do not arise in PDRs (see \S 3.3 and \S 4) the
similarity in the line ratios is 
probably a reflection of a similarity in the physical conditions:
relatively high temperatures and densities, optically thin CI
emission, and (marginally) optically thick CO emission. 

\subsection{The HCN line}

Our tentative detection of \hcnb43\ indicates a luminosity $L'$ (in K
km s$^{-1}$ pc$^2$) relative to the \cob43\ line of about 0.25, which
is similar to the \hcna10/\coonezero\ ratio seen in ultraluminous
galaxies (\cite{sol92b}). Except for $T\approx 60$ K, the lowest
allowable value, the maximum allowed densities in the CO-emitting
region ($n\sim 10^5\pcubcm$) are far below the effective
critical density required to thermalize the \hcnb43\ line [$n\approx
10^7\pcubcm$; even for the \hcna10\ line, the critical density is
$n\approx 10^5\pcubcm$ at these temperatures]. In this case the ratio
of HCN and CO luminosities is essentially determined by the
sub-thermal excitation of the HCN. (This is also true for models with
$T=60$ K unless the density is extremely high.) This is not a very
satisfactory explanation, however, as it requires fine-tuning of the
gas parameters to match the observed ratio, and does not explain why a
ratio of order 0.2 is characteristic of ultraluminous galaxies. A
similar constraint arises from the failure of Wilner \etal (1995) to
detect the HCO$^+$ $(4-3)$ line, implying that most of the gas is not
at densities above the effective critical density for this transition,
$n_{\rm cr}\approx 2\times 10^6\pcubcm$.

It is more probable that this ratio reflects the fraction of the
CO-emitting region which has a density high enough to collisionally
excite the HCN line. Solomon, Downes, \& Radford (1992b) argue that
only star-forming regions in molecular clouds have such high gas
densities, and that therefore ultraluminous galaxies are powered by
star formation. However, this argument overlooks the fact that in
galaxies with powerful active nuclei, such as the Cloverleaf and
F10214+4724, there is a large minimum gas pressure which is enforced
by the radiation pressure from the AGN.\footnote{
Here we assume the gas communicates the radiation pressure through
much of the molecular column in directly illuminated clouds.  This 
wouldn't necessarily be true with a highly
imhomogenous medium with an effective covering factor far above unity.}

Assuming that the incident flux from the AGN is absorbed in a region
which is thin compared to the thickness of a cloud, then in
equilibrium the cloud pressure must at minimum equal the absorbed radiation
pressure. In terms of $\tilde{P}=nT$, this constraint can be written
\be
\tilde{P}_{\rm min}=2\times 10^{10}{L_{44}\over \rpcsq}\left({\fabs\over
0.1}\right)\;{\rm cm^{-3}\ K}
\ee
where the bolometric luminosity of the AGN is $L_{\rm
bol}=10^{44}L_{44}$ erg s$^{-1}$, the distance of the cloud from the
AGN is \rpc\ parsecs, and the fraction of the incident luminosity
which is absorbed is \fabs. In terms of the gas density rather than
$\tilde{P}$, equation (1) is
\be
n_{\rm min}=2\times 10^8{L_{44}\over \rpcsq}\left({\fabs\over
0.1}\right)T_2^{-1}\;{\rm cm^{-3}}
\ee
where the gas temperature $T=100 T_2$ K. The bolometric luminosity of
the Cloverleaf is $L_{\rm bol}\approx 10^{14} m^{-1} h^{-2}\lsol$, where
$m$ is the magnification factor. Modelling of the Cloverleaf images
suggests that $m\sim 10$ (\cite{kay90}) but
this is uncertain. Equation (2)
predicts that the density will exceed $10^5\pcubcm$ [enough to
collisionally excite the \hcna10\ line] within a radius 
\be
r=8.8\times 10^2 \left({m\over 10}\right)^{-1/2} T_2^{-1/2}
\left({\fabs\over 0.1}\right)^{1/2}\;{\rm pc.}
\ee 
For $n\gtrapprox 10^7\pcubcm$, needed to excite the \hcnb43\ line, the
numerical coefficient is smaller by a factor of ten. The apparent line
fluxes for transitions which require very high densities
to excite are likely to be significantly affected by differential
magnification, as they will arise preferentially at small radii. 

An alternative possibility which may be generically important in
infrared-luminous galaxies -- either starbursts or AGNs -- is that the
HCN levels are not 
collisionally excited, but instead are populated through absorption of
$14\mu$m photons in the degenerate bending mode (e.g.,
\cite{aa94}). For optically thin HCN emission (no radiative
trapping), the criterion for radiative pumping of the \hcnb43\ line to
be significant is that the dust temperature fulfills the condition
\be
T_d\gtrapprox 1025[\ln(487 W+1)]^{-1}\;{\rm K}
\ee
where the continuum has been assumed to be a Planck
function multiplied by a dilution/optical depth factor $W$. For $W=1$,
equation (4) requires $T_d\gtrapprox 166$ K; for $W=0.1$,
$T_d\gtrapprox 262$ K. Substantial amounts of dust at these
temperatures are undoubtedly present in the Cloverleaf
(\cite{bar95}). We note that these two mechanisms predict rather
different size scales for the HCN emission; from equation (3),
radiation pressure will only keep the gas density high enough to
collisionally excite the \hcnb43\ line to a radius of 100 pc, whereas
for a bolometric luminosity of $10^{13}\lsol$, the dust temperature
will be above 250 K to more than 300 pc distance. We might also expect
that in the former case, the HCN and CO line profiles could differ
significantly; testing this will require much higher signal-to-noise
observations than are presently available.

\section{Discussion and Conclusions}

The presence of a luminous active nucleus appears to have profoundly
influenced the state of the interstellar medium in the Cloverleaf. The
high line luminosities, the high ratio of CI to CO line intensities,
and the prominent HCN emission may be the result of the impact of the
radiation from the AGN on the surrounding molecular clouds.

The mass of gas in the Cloverleaf can be estimated from both the CO
line observations and the CI 492 GHz line. The total molecular gas
mass can be derived from an observed CO line luminosity and our
detailed model results as \be M_\htwo({\rm CO}) = L_{\rm CO}(M_{\rm
cl}/ L_{\rm cl}) \approx 215 L_{\rm CO}{10^{-17}(\nco/\dv)\over
(y_{\rm CO}/3\times 10^{-5})}(\nu I_\nu)^{-1} \msol \ee where
\nco/\dv\ is in \psqcm\ km$^{-1}$ s, $\nu$ is the rest frequency of
the transition, $I_\nu$ is the intensity at line center emerging from
a cloud, and $y_{\rm CO}$ is the CO abundance by number relative to
total hydrogen; $M_{\rm cl}/ L_{\rm cl}$ is the mass-to-CO luminosity
ratio of the clouds from our models. Using the observed luminosity of
the \coa32\ line of $L_{\rm CO}(3-2)=1.0\times 10^8 m^{-1}
h^{-2}\lsol$, we find a molecular gas mass $M_\htwo\approx
2\times10^{10} m^{-1} h^{-2}\msol$, essentially independent of the
choice of model, (except for the highest column density,
high-temperature models, where it can be as large as $M_\htwo\approx
6\times10^{10} m^{-1} h^{-2}\msol$). This is an order of magnitude
smaller than the value originally suggested by \cite{bar94}, due to
the considerably higher emissivity of warm CO of modest optical depth.

To estimate the gas mass from the CI emission, we assume that the 492
GHz line is optically thin. Over the allowed range of physical
conditions (see \S 3.2) approximately 40\% of the carbon atoms are in
the $J=1$ level of the ground $^3P$ state. Using an Einstein
$A$-coefficient of $7.9\times 10^{-8}$ s$^{-1}$ (\cite{nuss71}), we find
that the mass of hydrogen associated with the CI is then
\be 
M_{\rm HI}({\rm CI})\simeq 1400 {L_{492}\over (y_C/3\times
10^{-5})}\msol 
\ee
where the line luminosity $L_{492}$ is in \lsol\ and $y_C$ is the
carbon abundance by number. (In both equations [5] and [6] we have
normalized to a carbon abundance of $1/10$ solar; equation [5] assumes
all carbon is in CO.) The observed line luminosity $L_{492}=5.8\times
10^7 m^{-1} h^{-2}\lsol$, and so we obtain a gas mass $M_{\rm
HI}({\rm CI})\approx 8\times 10^{10} m^{-1} h^{-2}\msol$. 

If irradiation by the central source is important in pressurizing the
clouds, there will be associated dynamical effects, as the radiative
momentum imparted to the clouds will provide an outward acceleration.
Scoville \etal (1995) suggested that support of the molecular gas
against its own self-gravity by radiation pressure could be important
in F10214+4724, in order to reconcile the fact that their derived
molecular mass exceeds the dynamical mass (estimated from the measured
CO size and linewidth) by an order of magnitude. We
estimate $M_{\rm dyn}$ for the Cloverleaf from the upper limit to
the CO source size $r$ (see footnote 2) and the observed linewidths:
\be
M_{\rm dyn}\approx {r \Delta V_{\rm FWHM}^2\over G \sin^2 i}\approx
1.6\times 10^{11} m^{-1} h^{-1}\msol ,
\ee
where the inclination angle $i$ has been taken to be $45^{\circ}$.
Comparison with the above expressions for
the masses derived from CO and CI shows that the gas mass is in both cases
smaller than the dynamical mass   
(assuming, of course, that $r$ is not much smaller
than the current upper limit, and that the carbon abundance is not
much less than $1/10$ solar). Although radiation pressure support
could be important in the Cloverleaf, it is not at present necessary
to appeal to it.

The gas mass derived from the CI observations is $\sim 1.3 - 4$ 
times larger than the molecular gas mass estimated from CO; either the
hydrogen associated with the atomic carbon dominates the gas mass
(plausibly a result of the high X-ray ionization and heating rates
within a few kpc of the active nucleus: \cite{mht96}), or else the CI
emission is being differentially magnified by the lensing relative to
the CO. In either case, the gas mass is substantial, but not extreme,
especially as the total magnification $m$ may be a factor of ten. The
detectability of the Cloverleaf in CO and CI would appear to be the
result not of an enormous mass of gas, but rather the effects of
gravitational lensing and the rather unusual physical conditions (high
emissivities) in the gas, in consequence of the proximity of a
luminous active nucleus.

\acknowledgments 
We are grateful to the referee, David Wilner, for insightful comments,
and to the IRAM staff for excellent support during the
observations. Stephane Guilloteau helped produce the \cod76 spectrum
from PdBI.  PRM acknowledges support through the NASA Long Term
Astrophysics Program under grant NAGW-4454.

\clearpage
\vglue 1.0truecm
\centerline {\bf FIGURE CAPTIONS}

\figcaption{CO, CI, and HCN lines from the Cloverleaf quasar, shown
with Gaussian fits to the line profiles.  See Table 1 for line 
parameters derived from the fits.  The \cod76 parameters have  
been estimated by fixing the gaussian-fitted line width to the mean of the
other CO lines, as described in \S 2 of the text. \label{fig1}}

\figcaption{ Allowed regions of CO column density per unit linewidth
versus molecular hydrogen density for three temperatures, based on 
non-LTE escape probability modelling of the CO line ratios. See
text for discussion.  \label{fig2}}

\clearpage
 
\begin{deluxetable}{rcccccccc}
\scriptsize
\tablecaption{Millimeter Line Observations of the Cloverleaf.
\tablenotemark{a} \label{tbl-1}}
\tablehead{
\colhead{Line} & \colhead{$\nu_{\rm obs}$\tablenotemark{b}} & \colhead{$T_A^*$}
& \colhead{$S_{\nu}$} & 
\colhead{$\Delta V_{\rm FWHM}$} & \colhead{$V$\tablenotemark{c}}  
&\colhead{$I$} 
&\colhead{$L$~/$10^{8}$} 
&\colhead{$L'$~/$10^{10}$} \nl
\colhead{~~} & \colhead{(GHz)}   & \colhead{(mK)}   
& \colhead{(mJy)} &  \colhead{(\kms )} & \colhead{(\kms )}  
& \colhead{(Jy \kms )} 
& \colhead{($L_{\sun}$)} 
&\colhead{(K km s$^{-1}$ pc$^2$)}
}
\startdata
\coa32 & 97.199  & 4.2(0.3) & 25.8(1.6) & 
362(23) & $-29(12)$ & 9.9(0.6) & 1.8(0.1) & $13.4(0.8)$ \nl
\13coa32 & 92.924 & $<1.1$ & $<6.7$ & ... & ... 
& $<2.4$ & $<0.4$ & $< 3.6$ \nl
\cob43 & 129.576 & 7.8(0.3) & 52.7(2.0) & 375(16) & $+12(7)$
& 21.1(0.8) & 5.1(0.2) & $16.1(0.6)$ \nl
\coc54 & 161.964 & 7.2(0.4) & 55.6(3.2) & 398(25) & $+5(12)$
& 24.0(1.4) & 7.2(0.4) & $11.7(0.7)$ \nl
\cod76 & 226.721  & ... & 109(5) & ... &  $-5(13)$
& 47.3(2.2) & 18.3(0.9) & $10.9(0.5)$    \nl
\car10    & 138.351  & 1.1(0.1) & 7.7(0.8)  & 430(46) & $-47(24)$
& 3.6(0.4) & 0.9(0.1) & $2.4(0.3)$ \nl
\carb21    & 227.510  & $<2.2$   & $<22.2$   & ... & ...    
& $<8.5$ & $<3.6$ & $<2.1$  \nl
\hcnb43    & 99.631  & 0.9(0.2) & 5.5(1.1)  & 436(103) & $+49(40)$    
& 2.6(0.5) & 0.5(0.1) & $3.4(0.6)$ \nl
\enddata

 
\tablenotetext{a}{Line parameters derived from Gaussian profile fits.
Quoted uncertainties are statistical only.
For \cod76, the line strength was estimated from
measurements taken at the Plateau de Bure 
Interferometer assuming a line FWHM of 376 km s$^{-1}$, as 
described in \S 2 of the text.
Upper limits are 3$\sigma$, and were derived from Gaussian fits with
the linewidth and redshift fixed to the mean values for the
detected lines.  Line luminosities assume $H_0 = 75$ km s$^{-1}$
Mpc$^{-1}$ and $q_0 = 0.5$, and are {\it not} corrected for gravitational
magnification; see text.}
\tablenotetext{b}{Observed frequency of line
centroid.}  
\tablenotetext{c}{Velocities measured relative to $z =
2.5579$, which is the weighted average redshift of the four detected
CO lines.}
 
\end{deluxetable}


\clearpage

\begin{thebibliography}{}

\bibitem[Aalto \etal 1991]{aa91} Aalto, S., Black, J.H., Booth,
R.S., \& Johansson, L.E.B. 1991, A\&A 247, 291

\bibitem[Aalto \etal 1995]{aa95} Aalto, S., Booth, R.S., Black,
J.H., \& Johansson, L.E.B. 1995, A\&A 300, 369

\bibitem[Aalto \etal 1994]{aa94} Aalto, S., Booth, R.S., Black,
J.H., Koribalski, B., \& Wielebinski, R. 1994, A\&A 286, 365

\bibitem[]{} Alloin, D., Guilloteau, S., Barvainis, R., Antonucci,
R., \& Tacconi, L.\ 1997, in preparation.

\bibitem[Barvainis et al. 1992]{bar92} Barvainis, R., Antonucci, R.,
\& Coleman, P.\ 1992, ApJ, 399, L19

\bibitem[Barvainis et al. (1994)]{bar94} Barvainis, R., Tacconi, L., 
    Antonucci, R., Alloin, D., \& Coleman, P. 1994, Nature, 371, 586

\bibitem[Barvainis et al. 1995]{bar95} Barvainis, R., Antonucci, R.,
Hurt, T., Coleman, P., \& Reuter, H.-P.\ 1995, ApJ, 451, L9

\bibitem[Broadhurst \& Leh\'ar 1995]{bro95} Broadhurst, T., \& Leh\'ar,
J.\ 1995, ApJ, 450, L41

\bibitem[Brown \& Vanden Bout1992a]{bro92a} Brown, R.L., \& Vanden Bout,
    P.A.\ 1992a, ApJ, 397, L11 

\bibitem[Brown \& Vanden Bout1992b]{bro92b} Brown, R.L., \& Vanden Bout,
    P.A.\ 1992b, ApJ, 397, L19 

\bibitem[]{} Burke, B.F., Leh\'ar, J., \& Conner, S.R.\ 1991, in 
{\it Lecture Notes in Physics}, eds.\ R. Kayser, D.\ Schramm, \& N.\ Nieser

\bibitem[B\"uttgenbach et al. (1992)]{but92}B\"uttgenbach, T.H.,
Keene, J., Phillips, T.G., \& Walker, C.K. 1992, ApJ, 397, L15

\bibitem[]{} Goodrich, R.W., Miller, J.S., Martel, A., Cohen, M.H., Tran,
H.D., Ogle, P.M., \& Vermeulen, R.C.\ 1996, ApJ, 456, L9

\bibitem[]{} Granato, G.L., Danese, L., \& Franceschini, A.\ 1996, ApJ, 460, L11

\bibitem[Kayser et al. 1990]{kay90} Kayser, R.\ et al. 1990, ApJ, 364, 15

\bibitem[Maloney \& Wolfire 1996]{mal96} Maloney, P.R., \& Wolfire,
M.G.\ 1996, in CO: 25 Years of Millimeter-Wave Spectroscopy,
ed.~W.B. Latter, S.J.E. Radford, P.R. Jewell, J.G. Mangum, and
J. Bally (Dordrecht: Kluwer), 299.

\bibitem[Maloney, Hollenbach \& Tielens 1996]{mht96} Maloney, P.R.,
Hollenbach, D.J., \& Tielens, A.G.G.M.\ 1996, ApJ, 466, 561.

\bibitem[Nussbaumer 1971]{nuss71} Nussbaumer, H. 1971, ApJ 166, 411

\bibitem[Ohta et al. (1996)]{} Ohta, K., Yamada, T., Bakanishe, K, Kohno, K.,
Akiyama, M., \& Kawabe, R. 1996, Nature, 382, 426

\bibitem[Omont et al. (1996)]{} Omont, A., Petitjean, P., Guilloteau, S.,
McMahon, R.G., Solomon, P.M., \& P\'econtal, E.\ 1996, Nature, 382, 428

\bibitem[]{} Rowan-Robinson, M., et al. 1991, Nature, 351, 719

\bibitem[]{} Sanders, D.B., Soifer, B.T., Elias, J.H., Madore, B.F>, Matthews, 
K., Neugebauer, G., \& Scoville, N.Z.\ 1988, ApJ, 325, 74

\bibitem[Schilke et al. (1993)]{sch93} Schilke, P., Carlstrom, J.E.,
Keene, J., \& Phillips, T.G. 1993, ApJ, 417, L67

\bibitem[Scoville \etal 1995]{sco95} Scoville, N.Z., Yun, M.S., Brown,
R.L., \& Vanden Bout, P.A. 1995, ApJ, 449, L109

\bibitem[Solomon, Downes, \& Radford 1992a]{sol92a} Solomon, P.M., Downes,
    D., \& Radford, S.J.E.\ 1992a, ApJ, 398, L29

\bibitem[Solomon, Downes, \& Radford 1992b]{sol92b} Solomon, P.M., Downes,
    D., \& Radford, S.J.E.\ 1992b, ApJ, 387, L55

\bibitem[White et al. (1994)]{whi94} White, G.J., Ellison, B., Claude,
S., Dent, W.R.F., \& Matheson D.N. 1994, A\&A 284, L23

\bibitem[Wills et al. 1992]{wil92} Wills, B.J., Wills, D., Evans,
N.J., Natta, A., Thompson, K.L., Breger, M., \& Sitko, M.L. 1992, ApJ,
400, 96

\bibitem[]{} Wilner, D.J., Zhao, J.-H., \& Ho, P.T.P.\ 1995, ApJ, 452, L91

\bibitem[Wright et al. 1991]{wri91} Wright, E.L., et al. 1991, ApJ,
381, 200 


\end{thebibliography}
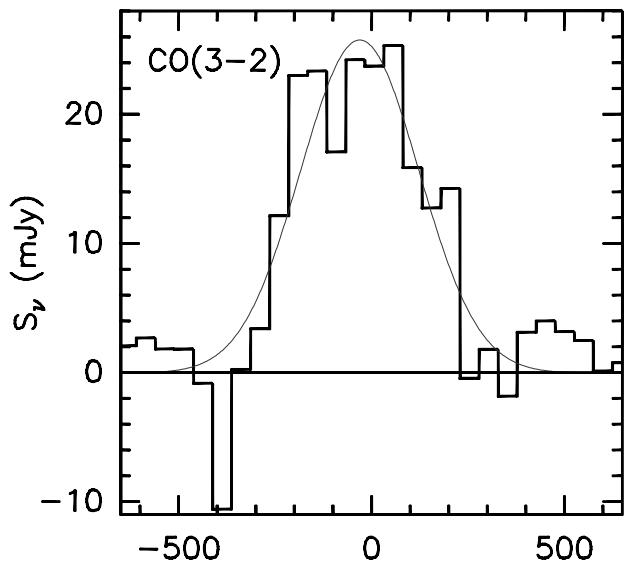
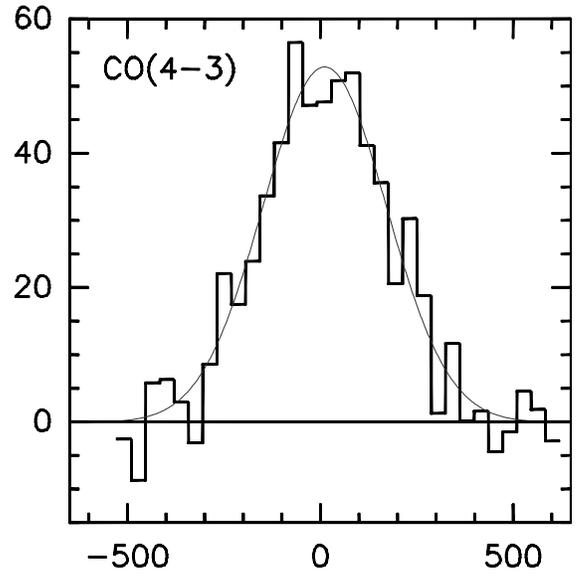
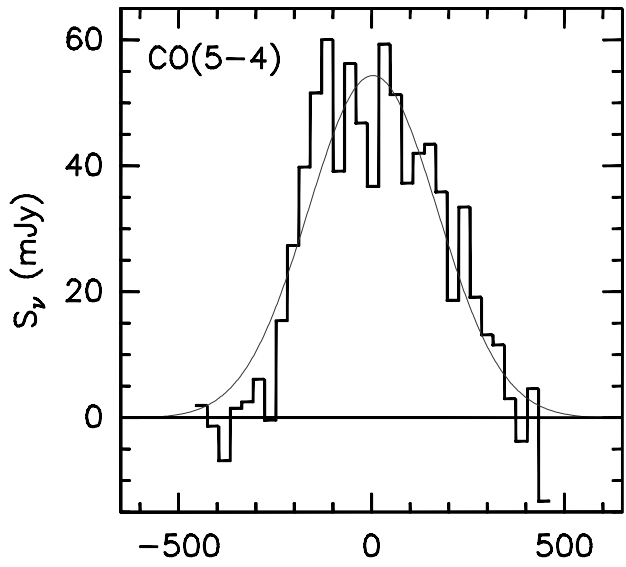
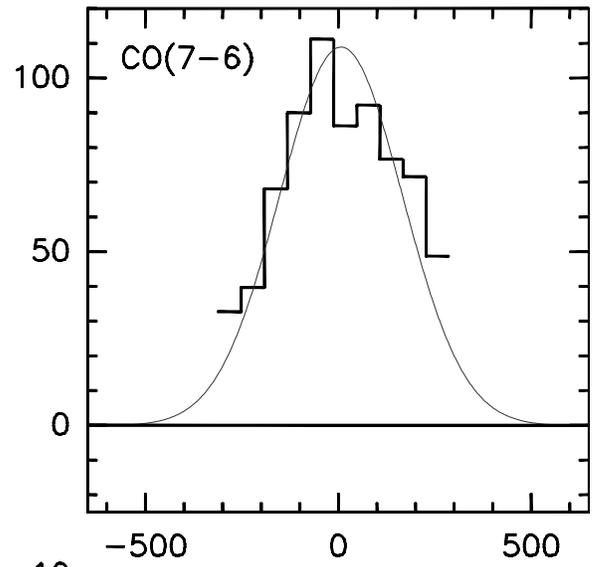
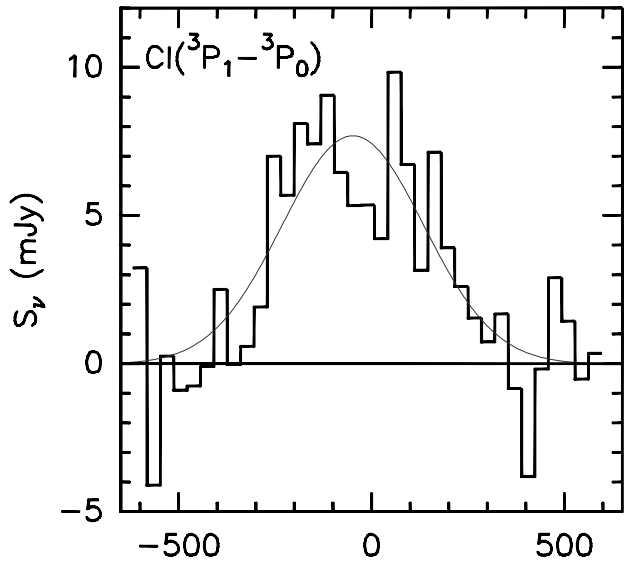
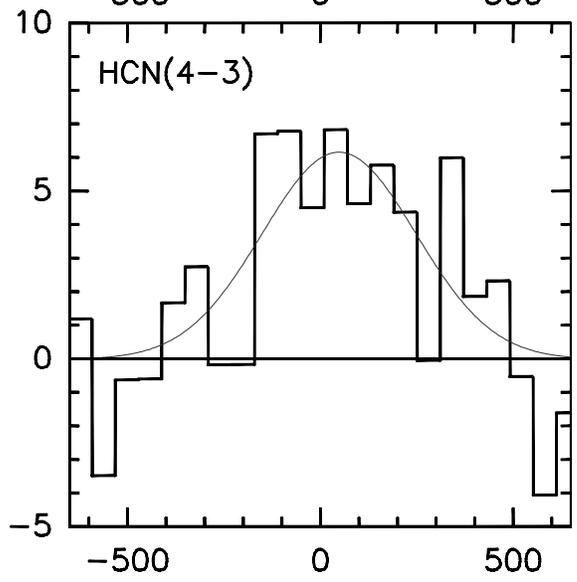
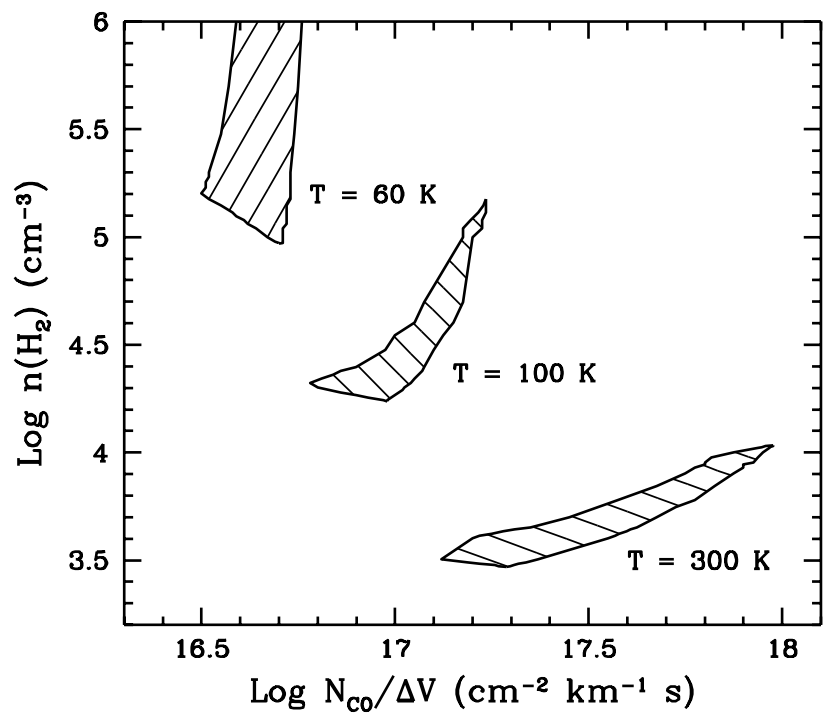
\end{document}